\documentclass[twocolumn,showpacs,nofootinbib,prd]{revtex4-2}
\usepackage{graphicx}
\usepackage{rotating}
\usepackage{color}
\usepackage{amssymb}

\newcommand{\tpz}{${}^{3\!}P_0$}
\newcommand{\bes}[1]{j^{#1}_{0}}
\newcommand{\han}[1]{h^{(1)#1}_{0}}

\newcommand{\tmat}[2]{T_{#1#2}}
\newcommand{\One}{1\;\!\!\!\!1}
\begin{document}
\title{Unitary model analysis of \boldmath$f_0(500)$ pole positions
by continuously varying \boldmath$m_\pi$: \\ comparison with discrete
lattice predictions}
\author{
George~Rupp}
\affiliation{
Center for Theoretical Particle Physics,
Instituto Superior T\'{e}cnico, Universidade de Lisboa,
P-1049-001 Lisboa, Portugal
}

\begin{abstract}
Resonance, bound-state, and virtual-state pole positions of the $f_0(500)$
scalar meson are computed as a continuous function of pion mass in the
framework of a unitarized and analytic coupled-channel model for scalar mesons,
described as dynamical quark-antiquark states. The $f_0(500)$ is modeled with
both light and strange $q\bar{q}$ seeds, mixing with each other through the
common $S$-wave $\pi\pi$, $K\bar{K}$, and $\eta\eta$ meson-meson decay
channels. The few model parameters are fitted to experimental $S$-wave $\pi\pi$
phase shifts up to 1~GeV. In the case of the physical $\pi^\pm$ mass of
139.57~MeV, resonance poles at $(460-i222)$~MeV and $(978-i37.2)$~MeV are found
for the $f_0(500)$ and $f_0(980)$, respectively. Resonance, bound-state, and
virtual-state pole trajectories are computed and plotted as a function of
pion masses up to 500~MeV, both in the complex-energy and complex-momentum
planes. The results are discussed and compared to the most advanced
lattice QCD computations employing interpolators that correspond to the
$q\bar{q}$ and meson-meson channels in the present model, that is, for a few
discrete values of the unphysical pion mass in those lattice calculations.
\end{abstract}

\maketitle

\section{Introduction}
\label{intro}
The light scalar mesons $f_0(500)$ (alias $\sigma$), $f_0(980)$,
$K_0^\star(700)$ (alias $\kappa$), and $a_0(980)$ \cite{PDG22} have long
eluded experimentalists as well as theorists (see Ref.~\cite{RB18} for a
minireview). On the one hand, the very existence of especially the extremely
broad $\sigma$ and $\kappa$ as genuine mesons has for many years
been questioned, due to the difficulty to identify unmistakable resonance
signals in scattering and production experiments. On the other hand, all four
scalars have been highly problematic to understand in the context of
conventional quark models, in view of their seemingly much too low masses as
$P$-wave quark-antiquark states. A very original and relatively successful
solution was proposed by R.~L.~Jaffe \cite{Jaffe77} in 1977, suggesting that
the above four scalars are $qq\bar{q}\bar{q}$ (tetraquark) states rather than
regular $q\bar{q}$ mesons. Owing to a very large and attractive 
color-spin interaction for the ground-state $S$-wave tetraquarks,
their masses come out several hundreds of MeV below the typical mass range of
1.3--1.5~GeV for $P$-wave mesons, thus allowing to predict much more reasonable
masses, albeit still somewhat on the high side \cite{Jaffe77,PDG22}.
Nevertheless, Jaffe himself admitted \cite{Jaffe77,RB17} that mass
calculations of mesons whose very large widths are ignored must not be taken
too literally and an accuracy as usually obtained for regular hadrons
should not be expected. He also observed \cite{Jaffe77,RB17} that such
$q^2\bar{q}^2$ systems can just fall apart into two light mesons, not
requiring the creation of a new $q\bar{q}$ pair like in the case of normal
mesons, thus being processes of order $N_c^0$ instead of $N_c^{-1}$.

However, a few years later, my co-authors and I found \cite{Eef83,Eef84} a
$\sigma$-meson resonance pole in the correct mass ballpark and with a
realistic width, in the framework of a unitarized coupled-channel model
\cite{Beveren83} for confined $q\bar{q}$ systems interacting with two-meson
states. This ``unquenched'' quark model had just been developed to describe
both light and heavy vector as well as pseudoscalar mesons by including the
nonperturbative dynamical effects of strong decay.
This $\sigma$ pole turned out to be of a dynamical origin, i.e., arising from
the $\pi\pi$ scattering continuum for increasing decay strength and not directly
linked to an intrinsic quark-model state. An equally dynamical $f_0(980)$ pole
appeared as well, just like the $\sigma(500)$ with a reasonable mass
and width, besides regular $f_0(1370)$ and $f_0(1500)$ \cite{PDG22} scalar
resonance poles predicted in mainstream quark models. Moreover, the gross
behavior of the $S$-wave $\pi\pi$ scattering phase
shifts was also automatically reproduced \cite{Eef83}. In
Ref.~\cite{Beveren86}, this model calculation was extended to the other light
scalars as well, resulting in predicted resonance pole positions for
the $\sigma$, $\kappa$, $f_0(980)$, and $a_0(980)$ that are still within
current PDG limits \cite{PDG22}, using the very same parameter values as those
found in the fit carried out in Ref.~\cite{Beveren83}.

When the four light scalars had finally been confirmed experimentally and
lattice QCD (LQCD) had made sufficient progress so as to be capable of
carrying out some reliable simulations of meson resonances,
first results on the light scalars started to appear in the literature
(see the review paper Ref.~\cite{Briceno18-1}). In particular,
in Ref.~\cite{Briceno17} an LQCD computation of isoscalar
$S$-wave $\pi\pi$ phase shifts was done for the two unphysical pion masses of
391~MeV and 236~MeV, while searching for bound states and/or
resonance poles. The simulation employed single-meson $u\bar{u}\!+\!d\bar{d}$
and $s\bar{s}$ interpolating fields as well as $\pi\pi$ and $K\bar{K}$
two-meson interpolators. As a result, a $\pi\pi$ bound state was found 
at $758\,(4)$~MeV for the heavier pion mass, whereas $\sigma$-like
resonance pole positions were extracted for the lighter pion, using a
variety of parametrizations, albeit widely
spread out in energy and with large to very large error bars.
A year later, a more extensive and detailed lattice analysis \cite{Briceno18-2}
of not only the light scalars, but also the lowest isoscalar tensor mesons
$f_2(1270)$ and $f_2^\prime(1525)$. In the case of the $\sigma$ and
$f_0(980)$, an $\eta\eta$ interpolating field was included as well. In this
simulation, only $m_\pi\!=\!391$~MeV was considered, resulting in a
$\sigma$-type bound state slightly lighter than in Ref.~\cite{Briceno17},
viz.\ at $745\,(5)$~MeV.

In the present Letter, I will present a unitarized coupled-channel model 
for the $\sigma$ resonance, in the spirit of Ref.~\cite{Beveren86} yet
formulated in momentum space following the Resonance Spectrum Expansion
(RSE) \cite{BR03,BR09}. The goal is to study in detail the behavior of the
$\sigma$ pole, not only for the two unphysical pion masses of 391~MeV
\cite{Briceno17,Briceno18-2} and 236~MeV \cite{Briceno17}, but rather as a 
continuous function of $m_\pi$ over a wide range of values starting at the
pion's physical mass. As the model is manifestly unitary and analytic,
it may shed some light on aspects of the parametrizations employed in
Refs.~\cite{Briceno17,Briceno18-2}. Since after I concluded the
present study my attention was drawn to two very recent LQCD papers involving
authors of Refs.~\cite{Briceno17,Briceno18-2} that present results for two
intermediate pion masses \cite{Rodas23a} and extracted $\sigma$ pole positions
further restricted by dispersive methods \cite{Rodas23b}, I shall briefly
discuss these papers below as well.

Another purpose of the present paper is to see how precisely the
$\sigma$ resonance pole for varying $m_\pi$ is connected not only to
a bound state but also to possible virtual states. This will furthermore serve
to explicitly check a claim made in the dispersive analysis of
Ref.~\cite{Gao22} about a virtual $\sigma$ pole not far from the bound-state
pole at $758\,(4)$ as found in Ref.~\cite{Briceno17} for $m_\pi\!=\!391$~MeV.
A Comment on this claim was already published in Ref.~\cite{BR23}, in the
context of a model \cite{BR21} very similar to the present one, but vaster in
scope so as to also investigate the $f_0(980)$ and $a_0(980)$, as well as the
standard quark-model scalar mesons $f_0(1370)$ and $a_0(1450)$.

This Letter is organized as follows. In Sec.~\ref{rsem} the here employed RSE
model will be described in detail, including a closed-form expression for the
multichannel $S$-matrix, and a fit to $S$-wave $\pi\pi$ phase shifts will be
carried out. In Section~\ref{trajectories} I will present several figures of
the resulting $\sigma$ resonance, bound-state, and virtual-state trajectories 
as a continuous function of pion mass. Finally, Sec.~\ref{conclusions}
will be devoted to some discussion and conclusions in connection with the
mentioned LQCD results.
\begin{figure*}[ht]
\centering
\includegraphics[trim = 43mm 212mm 7mm 45mm,clip,width=17cm,angle=0]
{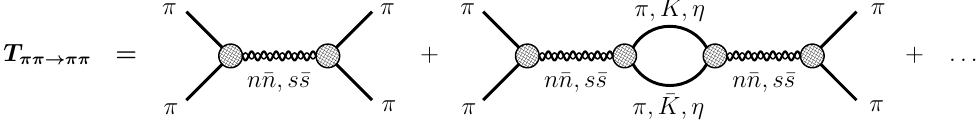}
\caption{Graphical representation of the $\pi\pi\to\pi\pi$ element of the 
$3\times3$ $T$-matrix, with $\pi\pi$, $K\bar{K}$, and $\eta\eta$ loops between
the $n\bar{n}$ and $s\bar{s}$ propagators in the intermediate state. The dots
stand for all higher-order terms in the $s$-channel bubble sum.}
\label{sigma}
\end{figure*}
\section{RSE Model for an isoscalar scalar}
\label{rsem}
The RSE model for nonexotic meson spectroscopy is based on an $s$-channel
propagator of a bare meson and its excitations, which couple in the 
intermediate state to an incoming pair of mesons and then to an outgoing meson
pair that may be different from the incoming one. Note that this
model is not suited to describe meson-meson (MM) scattering for quantum numbers
not supported by $q\bar{q}$ states, as e.g.\ $I=2$. Also, crossing-symmetry
constraints are not explicitly imposed. However, it has been argued by
different authors that nonexotic meson resonances are dominated by several
$s$-channel exchanges, which according to duality also account for some
$t$- and $u$-channel phenomena (see Ref.~\cite{BR23} for references). The RSE
bubble sum for the $\pi\pi\to\pi\pi$ $T$-matrix element is depicted in
FIG.~\ref{sigma}, where we take towers of bare
$n\bar{n}\equiv(u\bar{u}+d\bar{d})/\sqrt{2}$ and
$s\bar{s}$ states coupling to the asymptotic two-meson channels $\pi\pi$,
$K\bar{K}$, and $\eta\eta$. The Born term in the figure stands for two incoming
pions, which at the initial vertex couple to intermediate $n\bar{n}$ and
$s\bar{s}$ states with all their excitations --- represented by the wiggly line
--- via one $q\bar{q}$ annihilation. After propagation of these states, a new
$q\bar{q}$ pair is created at the second vertex, giving rise to two pions again
in the final state. Note that we assume $q\bar{q}$ ($n\bar{n}$ or
$s\bar{s}$) creation and annihilation to take place according to the
empirically successful \tpz\ model \cite{Micu69}, that is, with vacuum quantum
numbers $J^{PC}\!=\!0^{++}$. At first sight it may seem odd to couple
$s\bar{s}$ states to pions, but the next-order diagram in the figure shows that
the $K\bar{K}$ and $\eta\eta$ loops in the intermediate state will inevitably
lead to a mixing of the bare $n\bar{n}$ and $s\bar{s}$ states, thus allowing
both mixed states to couple to $\pi\pi$. Clearly, the whole bubble sum can be
easily summed up algebraically. In other words, since the effective interaction
represented by the Born term in the figure is separable, just like similar
diagrams with $K\bar{K}$ and/or $\eta\eta$ pairs in initial and/or final state,
the complete three-channel $T$-matrix can be straightforwardly solved in
closed form.

The full expression for the effective interaction, symbolized by the Born
term for the $\pi\pi\to\pi\pi$ process in FIG.~\ref{sigma}, reads explicitly
(also see Ref.~\cite{BR21})
\begin{equation}
V_{ij}(p_i,p'_j;E)  =  
\lambda^2j^i_{0}(p_ia)\,\mathcal{R}_{ij}(E)\,j^j_{0}(p'_ja) \;,
\label{veff}
\end{equation}
\begin{equation}
\mathcal{R}_{ij}(E)=\sum_{\alpha=1}^{2}\sum_{n=0}^{\infty}
\frac{g^i_{(\alpha,n)}g^j_{(\alpha,n)}}{E-E_n^{(\alpha)}}\;,
\label{rse}
\end{equation}
where the RSE propagator $\mathcal{R}$ contains an infinite tower of bare
$n\bar{n}$ and $s\bar{s}$ states with quantum numbers $J^{PC}=0^{++}$\,
corresponding to the discrete scalar spectrum of an in principle arbitrary
confining potential. Also, $E_n^{(\alpha)}$ is the energy level of the $n$-th
recurrence in the $\alpha$-th $q\bar{q}$ channel, with $\alpha=1$
referring to $n\bar{n}$ and $\alpha=2$ to $s\bar{s}$, while $g^i_{(\alpha,n)}$
is the corresponding coupling to the $i$-th MM channel. Furthermore, in
Eq.~(\ref{veff}), $\lambda$ is the overall coupling constant for \tpz\ decay,
and $\bes{i}(p_i)$ and $p_i$ are the zeroth order ($L=0$) spherical Bessel
function and the (relativistically defined) off-energy-shell relative momentum
in MM channel $i$, respectively. The spherical Bessel function
originates in our sharp string-breaking picture of OZI-allowed decay at a
certain radius $a$, being the Fourier transform of a spherical delta-shell at
$r=a$. Such a picture is supported by LQCD simulations \cite{Bali05}.
The channel couplings $g^i_{(\alpha,n)}$ in Eq.~(\ref{rse}) are computed
following the formalism developed in Ref.~\cite{B82}, namely from
overlaps of harmonic-oscillator (HO) wave functions for the original
$q\bar{q}$ pair, the created \tpz\ pair, and the two $q\bar{q}$ states
corresponding to the outgoing two mesons. In most cases, this method produces
the same couplings for ground-state mesons as the usual point-particle
recoupling schemes of spin, isospin, and orbital angular momentum, but it also
provides a clear prescription for excited mesons, with the additional
advantage of always resulting in a finite number of nonvanishing couplings.
Because of their fast decrease for increasing radial quantum number $n$,
practical convergence of the infinite sum in Eq.~(\ref{rse}) is achieved by
truncating it after at most 20 terms. Moreover, the method guarantees
rigorous flavor symmetry for all nonexotic mesons, including the $\sigma$
(see discussion and examples in Refs.~\cite{BR99a,BR99b,BR99c}).

With the separable effective and energy-dependent MM potential in
Eqs.~(\ref{veff},\ref{rse}), the $3\times3$ fully off-energy-shell $T$-matrix
can be solved directly, yielding
\begin{eqnarray}
\lefteqn{\tmat{i}{j}(p_i,p'_j;E)=-2a\lambda^2\sqrt{\mu_ip_i}\,\bes{i}(p_ia)
\times} \nonumber \\
&&\hspace*{-1pt}\sum_{m=1}^{N}\mathcal{R}_{im}\left\{[\One-\Omega\,
\mathcal{R}]^{-1}\right\}_{\!mj}\bes{j}(p'_ja)\,\sqrt{\mu_jp'_j} \; ,
\label{tmat}
\end{eqnarray}
with the loop function
\begin{equation}
\Omega_{ij}(k_j)=-2ia\lambda^2\mu_jk_j\,\bes{j}(k_ja)\,\han{j}(k_ja)\,
\delta_{ij}\;, 
\label{omega}
\end{equation}
where $\han{j}(k_ja)$ is the spherical Hankel function of the first kind,
$k_j$ and $\mu_j$ are the on-shell relativistic relative momentum and reduced
mass in MM channel $j$, respectively, and the matrix $\mathcal{R}(E)$ is
given by Eq.~(\ref{rse}). Note that no regularisation is needed in this
model to all orders, since the Bessel functions at the vertices make the meson
loops finite. The manifestly analytic and unitary $S$-matrix is simply
given by
\begin{equation}
\mathcal{S} = \One + 2i \, \hat{T} \; ,
\label{smatrix}
\end{equation}
where $\hat{T}$ is the fully on-energy-shell submatrix of $T$, restricted
to the kinematically allowed MM channels just like $\mathcal{S}$.
From Eqs.~(\ref{veff}--\ref{smatrix}), the cotangent of the $S$-wave $\pi\pi$
phase shift can then be expressed as
\begin{equation}
\cot(\delta^{(0)}_{\pi\pi})\, = \, i \, + \, \frac{\eta^{(0)}_{\pi\pi}}
{\displaystyle\hat{T}_{\pi\pi\to\pi\pi}-i\,\frac{1-\eta^{(0)}_{\pi\pi}}{2}}\;,
\label{phase}
\end{equation}
where $\eta^{(0)}_{\pi\pi}$ is the inelasticity in the 
$\pi\pi\to\pi\pi$ channel, with $\eta^{(0)}_{\pi\pi}=1$ for $E<2m_K$ and
$|\mathcal{S}_{\pi\pi\to\pi\pi}|$ otherwise.

Before we now carry out a fit to the $S$-wave $\pi\pi$ phase shift, let us
introduce one more phenomenological degree of freedom, namely the intrinsic
mixing between the bare $n\bar{n}$ and $s\bar{s}$ states. It is true that the
common $K\bar{K}$ and $\eta\eta$ channels, via $s\bar{s}$ and/or $n\bar{n}$
creation, inexorably lead to such a mixing beyond the Born approximation, but
an additional mixing already at the quark level is perfectly possible, namely
via two gluons. Since low-energy QCD does not allow to rigorously compute this
mixing, we here introduce the corresponding angle $\Theta_S$ as a free fit
parameter, just like in Ref.~\cite{BR21}. The other parameters to be varied in
the fit are the overall coupling $\lambda$ and the decay radius $a$. The
original model parameters we keep exactly equal to their values as fixed in
Ref.~\cite{Beveren83} and then used in Ref.~\cite{Beveren86} as well as in all
posterior model calculations, both in coordinate-space and momentum-space
approaches. These fixed parameters are: the constituent quark masses 
$m_n=406$~MeV and $m_s=508$~MeV, besides the constant level splittings
$2\omega=380$~MeV and $\omega=190$~MeV between radial and orbital excitations,
respectively. The latter equidistant HO spectrum is clearly not a canonical
one in meson spectroscopy, but its choice is immaterial for the present study,
because the resulting lowest standard $n\bar{n}$ and $s\bar{s}$ states come
out at about 1.3~GeV and 1.5~GeV, respectively, very close to the values found
in mainstream quark models. For theoretical and empirical justifications of
a mass-dependent HO potential, see Refs.~\cite{Eef84,BR20}. The latter 
reference concerns a review paper that revisits many successful
meson-spectroscopy applications of such a potential in the framework of
unitarized quark models.

\begin{figure}[!t]
\centering
\includegraphics[trim = 43mm 95mm 20mm 30mm,clip,width=8cm,angle=0]
{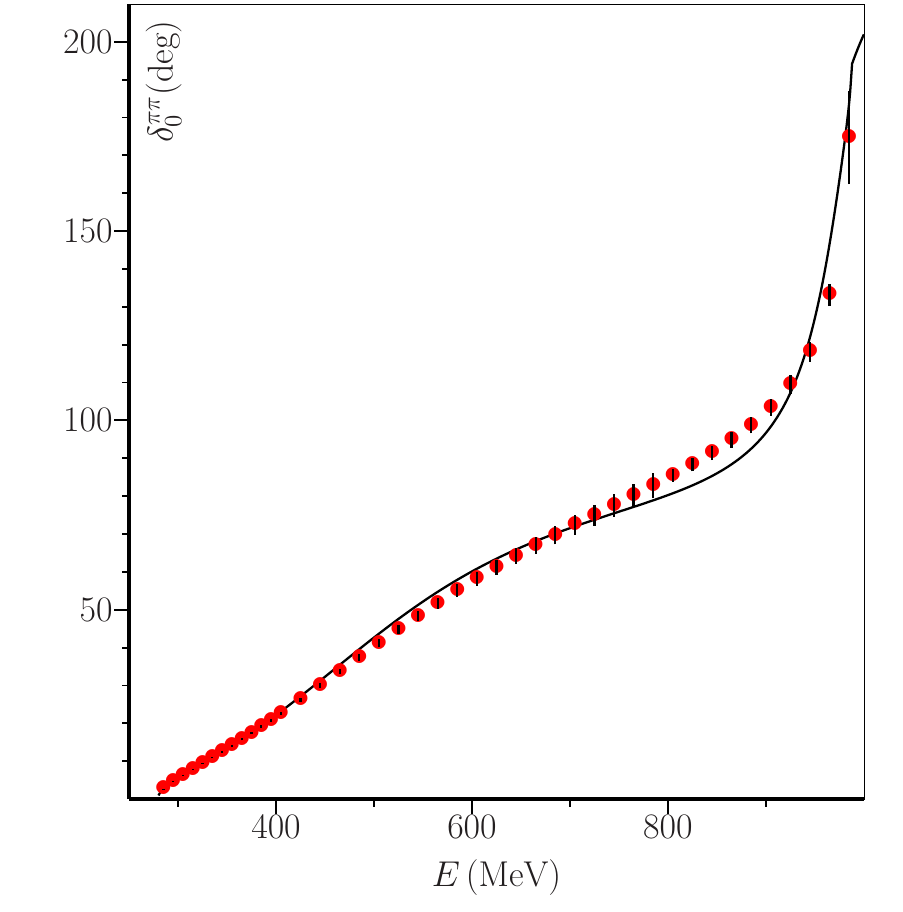} 
\caption{Model fit to $S$-wave $\pi\pi$ phase shifts from Ref.~\cite{Bugg06}.}
\label{pipi}
\end{figure}
Now we are in a position to fit $\lambda$, $a$, and $\Theta_S$ to $S$-wave
$\pi\pi$ phase shifts from threshold up to 1~GeV collected and compiled by
D.~V.~Bugg \cite{Bugg06} from various sources (also see Ref.~\cite{BBKR06}).
The fitted parameter values are\footnote{
Note that in Ref.~\cite{BR21} a wrong dimension of GeV$^{-1/2}$ was
specified for $\lambda$. With $\lambda$ defined as in
Eqs.~(\ref{veff}--\ref{omega}), it is dimensionless.}
\begin{equation}
\lambda=3.673, \; a=3.314\,\mbox{GeV}^{-1}, \;
\Theta_S=7.515^\circ \; .
\label{par}
\end{equation}
These values of $\lambda$ and $a$ are close to those found in
Ref.~\cite{BR21}. But note that the fitted $\Theta_S$ is much smaller here
than in Ref.~\cite{BR21}, which is plausible, as in the latter paper a much
larger energy interval had to be accommodated. For the same reason, we now do
not impose an explicit extra damping of closed channels like in
Ref.~\cite{BR21}. Owing to this small value of $\Theta_S$, our bare $n\bar{n}$
and $s\bar{s}$ states are almost pure, with most of the mixing between the
$\sigma$ and $f_0(980)$ resulting from the common $K\bar{K}$ and $\eta\eta$
channels. The result of the fit to the $\pi\pi$ phases is shown in
FIG.~\ref{pipi}. The quality of the fit is satisfactory, considering the mere
three adjustable parameters. The kink in the model curve at about 990~MeV
is due to the opening of the $K\bar{K}$ channel, which introduces an $S$-wave
inelasticity in the $\pi\pi$ channel. Also note that the almost perfect
fit at low energies is reflected in the obtained isoscalar $S$-wave $\pi\pi$
scattering length $a_0^0=0.211\,m_\pi^{-1}$. Finally, we find the two
isoscalar scalar resonance poles on the second Riemann sheet (in MeV)
$\sigma(460-i222)$ and $f_0(978-i37.2)$.

\section{\boldmath$\sigma(500)$ poles as a function of $m_\pi$}
\label{trajectories}
As already mentioned in the Introduction, Ref.~\cite{Briceno18-2} has
studied the isoscalar scalar system in lattice simulations for two unphysical
pion masses, viz.\ $m_\pi=391$~MeV \cite{Briceno17,Briceno18-2} and
$m_\pi=236$~MeV \cite{Briceno17}. For the smaller pion mass, a resonance
pole could be extracted resembling the $\sigma(500)$, albeit with a too
large real part. However, this extraction is difficult and leads to widely 
scattered values for the real and imaginary part of the pole,
depending on the employed parametrization \cite{Briceno17}. Moreover, in order
to extrapolate these results towards the physical pion mass, different methods
may be used as well. I that spirit, we compute here $\sigma$ pole positions
as a continuous function of $m_\pi$, leading to bound-state, virtual-state,
and resonance trajectories, which we shall display next both in the complex
energy and momentum planes.

The numerical technique I use to search for real or complex poles in the
$\mathcal{S}$-matrix is by finding the zeroes in the modulus squared of the
determinant of the matrix $\One-\Omega\,\mathcal{R}$ in Eq.~(\ref{tmat}), just
like already done for the $\sigma(500)$ and $f_0(980)$ resonances
above. This is easy by employing the Fortran-coded minimization package
``MINUIT'' of the CERN Program Library \cite{minuit}.

\begin{figure}[!t]
\centering
\includegraphics[trim = 43mm 95mm 20mm 23mm,clip,width=8cm,angle=0]
{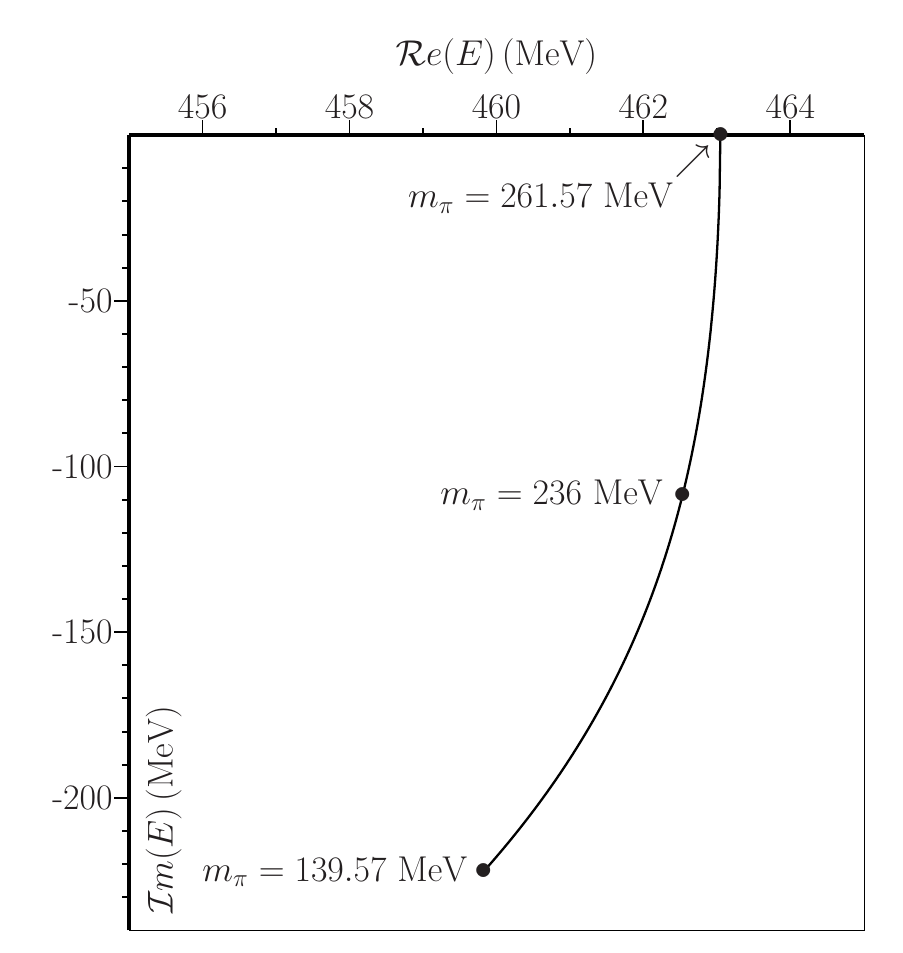} 
\caption{Resonance pole trajectory of $\sigma(500)$ in the complex
$E$ plane as a function of $m_\pi$.}
\label{rese}
\end{figure}

\begin{figure}[!t]
\centering
\includegraphics[trim = 43mm 95mm 20mm 30mm,clip,width=8cm,angle=0]
{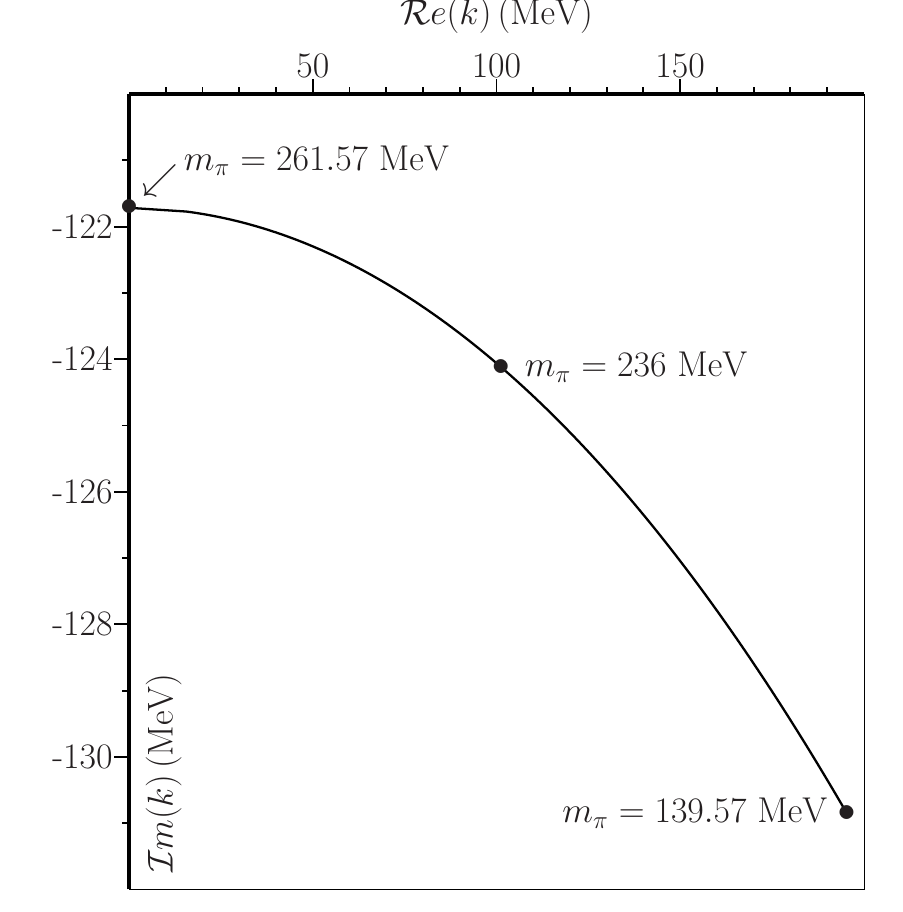} 
\caption{Resonance pole trajectory of $\sigma(500)$ in the complex
$k$ plane as a function of $m_\pi$, for $\mathcal{R}e(k)>0$.}
\label{resk}
\end{figure}

\begin{figure}[!b]
\centering
\includegraphics[trim = 43mm 95mm 20mm 30mm,clip,width=8cm,angle=0]
{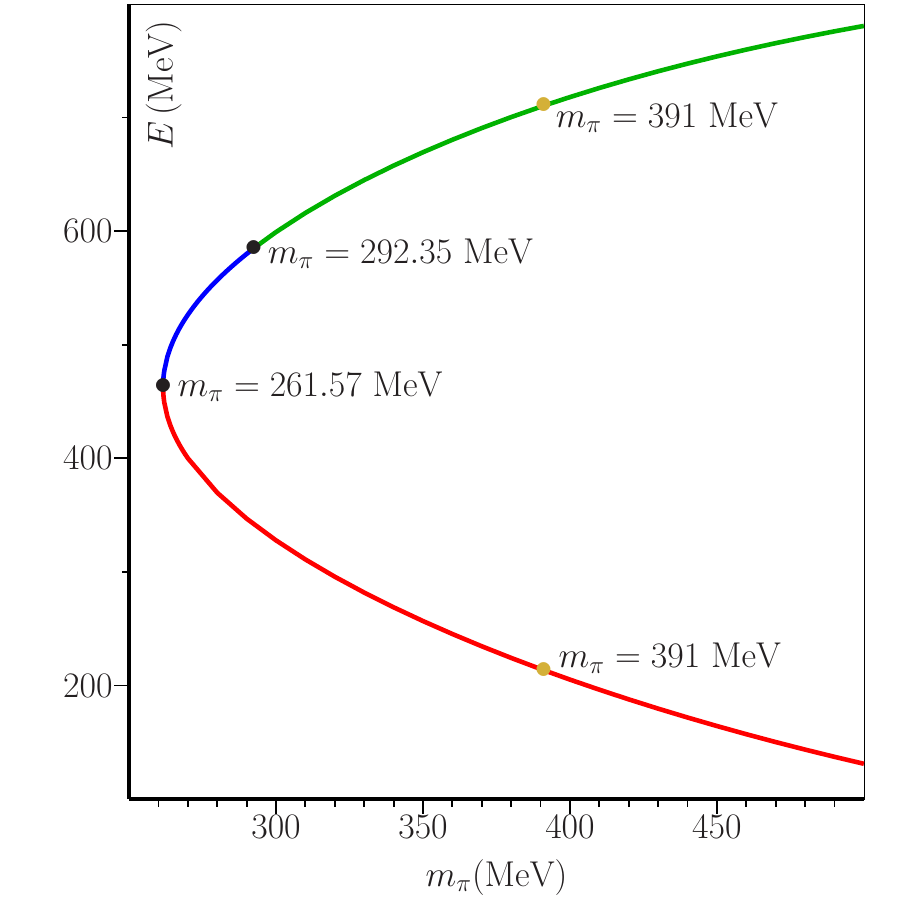} 
\caption{Bound-state and virtual-state pole trajectories of real $\sigma(500)$
energy as a function of $m_\pi$; green (upper section): bound state,
blue (middle section): first virtual state, red (lower section):
second virtual state.}
\label{bve}
\end{figure}

\begin{figure}[t]
\centering
\includegraphics[trim = 43mm 95mm 20mm 30mm,clip,width=8cm,angle=0]
{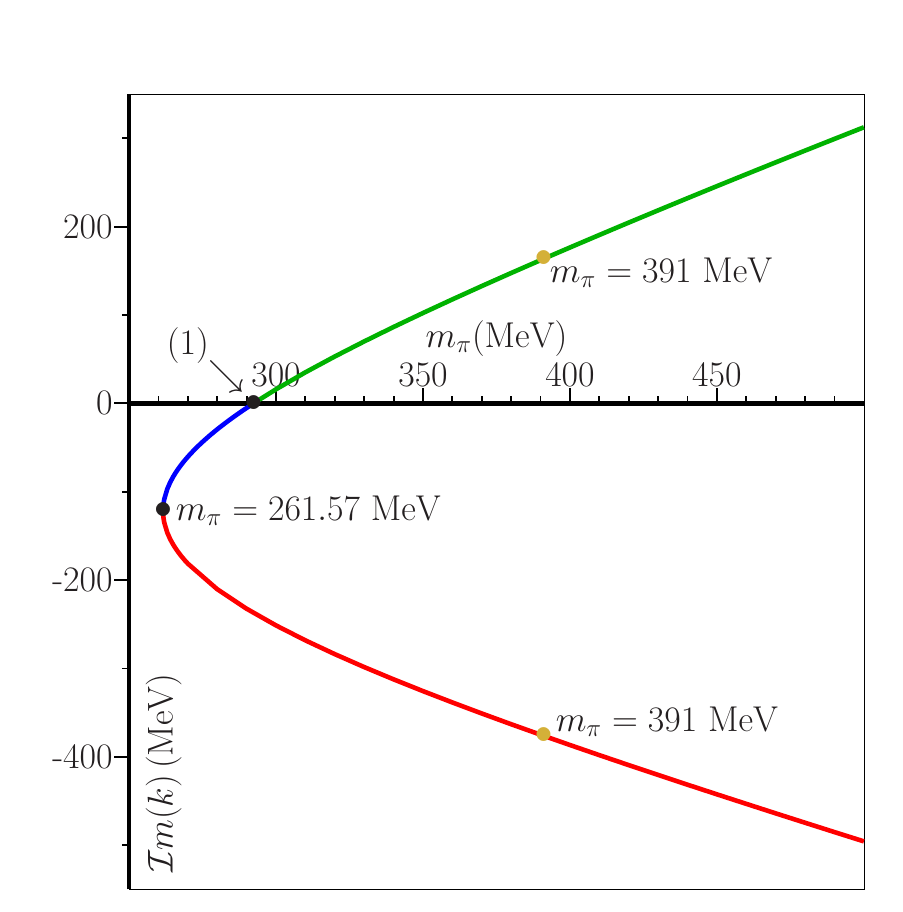} 
\caption{Bound-state and virtual-state pole trajectories of imaginary
$\sigma(500)$ momentum as a function of $m_\pi$; green (upper section):
bound state, blue (middle section): first virtual state, red (lower section):
second virtual state. Point (1): $m_\pi=292.35$ MeV.}
\label{bvk}
\end{figure}

In FIG.~\ref{rese} the resonance pole trajectory of the $\sigma(500)$ in the
complex $E$ plane is shown for a pion mass ranging from its physical value
to $m_\pi=261.57$~MeV, where the pole reaches the real axis and splits
into a pair of virtual-state poles, i.e., staying in the second Riemann sheet.
Note that for $m_\pi > 231.21$~MeV we are dealing with a typical subthreshold
$S$-wave resonance, just like in the case of making the resonance pole move to
the real axis by increasing the overall coupling $\lambda$ (see FIG.~2 in
Ref.~\cite{BR23}). What is quite remarkable is the enormous stability of the
real part of the resonance pole, increasing by only about 3~MeV 
while the imaginary part runs from -222~MeV to zero. Figure~\ref{rese} also
indicates the resonance pole for the pion mass of 236~MeV employed in
Ref.~\cite{Briceno17}. The equivalent pole trajectory in the complex $k$ plane
is displayed in FIG.~\ref{resk}, where the mirror-image trajectory with
$\mathcal{R}e(k)<0$, required for unitarity \cite{BR20}, is not shown. The
latter two poles meet on the negative imaginary $k$ axis for
$m_\pi=261.57$~MeV, indeed confirming that we are dealing now with virtual
poles. This typical $S$-wave behavior of $S$-matrix poles is further clarified
by allowing $m_\pi$ to become even larger. So in FIG.~\ref{bve} the real
trajectories of bound-state and virtual-state poles are displayed by plotting
the corresponding energies directly as a double-valued function of $m_\pi$, up
to $m_\pi=500$~MeV. First of all, a $\sigma(500)$ bound state is found at
$E=710.3$~MeV for the pion mass of 391~MeV used in
Refs.~\cite{Briceno17,Briceno18-2}. We also see that, for all values of
$m_\pi>261.57$~MeV, there are two real poles, i.e., one bound state plus one
virtual state or two virtual states, the latter meeting at 261.57~MeV to
transform into one resonance for smaller values of $m_\pi$. Also this behavior
is qualitatively the same as that observed in the case of pole trajectories as
a function of $\lambda$ \cite{BR21,BR23}. Moreover, FIG.~\ref{bve} shows that 
in the bound-state situation, there is only a very far-away virtual pole,
contrary to the claim in Ref.~\cite{Gao22}, as argued in Ref.~\cite{BR23} as
well. Finally, FIG.~\ref{bvk} displays the
corresponding (imaginary) momentum trajectory of bound and virtual states.

\section{Discussion and Conclusions}
\label{conclusions}
Before comparing the above model results to recent LQCD predictions, let me
mention two related works from about a decade ago. In
Ref.~\cite{Hanhart14}, it was shown that the trajectories of poles as a
function of some strength parameter and coupling to continuum $P$-waves or
higher are qualitatively the same, irrespective of the employed dynamics. On
the other hand, in the case of $S$-wave pole trajectories it was suggested
that important information on the internal structure of a resonance can be
revealed. As for Ref.~\cite{Albaladejo12}, the authors made use of unitary
chiral perturbation theory to study the pion-mass dependence of the size of
the $\sigma(500)$, then on the verge of being renamed from $f_0(600)$ to
$f_0(500)$ \cite{RB18}. They concluded that for a large pion
mass, of the order of 400~MeV, the picture of a molecular-type, spread-out
$\pi\pi$ state appears to be suitable. This would indeed resemble
somewhat the relatively small $\pi\pi$ binding for $m_\pi=391$~MeV, as later
found on the lattice in Refs.~\cite{Briceno17,Briceno18-2,Rodas23a,Rodas23b}.
However, for the physical pion mass they obtained a small $\sigma(500)$ size
suggesting a compact object, for which the authors considered a four-quark
picture more appropriate. I disagree with the latter assessment, based on the
above model results, as well as the referred lattice computations,
in which no four-quark interpolators were employed.

We will now compare our continuous $\sigma(500)$ trajectories with the
discrete $\sigma$ lattice pole positions for $m_\pi=391$~MeV
\cite{Briceno17,Briceno18-2} and $m_\pi=236$~MeV \cite{Briceno17}. Starting
with the bound-state case, we recall the model prediction 710.3~MeV, whereas
the lattice yielded $758\,(4)$~MeV \cite{Briceno17} and $745\,(5)$~MeV
\cite{Briceno18-2}, for $m_\pi=391$~MeV. This discrepancy is probably due to
a difference in the handling of the $K\bar{K}$ and $\eta\eta$ channels. First
of all, in Ref.~\cite{Briceno18-2} Kaon and $\eta$ masses were
employed of 549~MeV and 587~MeV, respectively, instead of the physical ones,
as a result of the chosen light and strange quark masses.
Although such an increased $K$ mass was not mentioned in
Ref.~\cite{Briceno17}, probably the same or a similar value was taken. The
small difference in binding energy of 13~MeV between the two simulations
in Refs.~\cite{Briceno17,Briceno18-2} appears to be indeed due to the inclusion
of the $\eta\eta$ channel in Ref.~\cite{Briceno18-2}, by providing some extra
attraction owing to the latter kinematically closed channel. If we
also increase the Kaon and $\eta$ masses in the model as in
Ref.~\cite{Briceno18-2}, the $\sigma$ mass comes out at 718~MeV. Now we should
recall that, in the more general model of Ref.~\cite{BR21}, a phenomenological
damping of closed channels was introduced in order to reduce their influence
far underneath the thresholds, which becomes necessary when simultaneously
fitting data over a very wide energy range. A similar suppression of closed 
channels was successfully used in Ref.~\cite{BBKR06}, dealing with the complete
scalar nonet. As mentioned in Ref.~\cite{BR23}, a $\sigma$ bound state of
760~MeV is found in the full model of Ref.~\cite{BR21}, for a pion mass of
391~MeV. If we here use the same subthreshold suppression of the $K\bar{K}$ and
$\eta\eta$ channels as in Ref.~\cite{BR21}, the bound state comes out at
752~MeV.

In the very recent LQCD paper of Ref.~\cite{Rodas23a}, further information
on the quark-mass dependence was obtained by carrying out computations of the
$\sigma$ pole at two intermediate pion masses, viz.\ $m_\pi=283$~MeV and
$m_\pi=330$~MeV. The authors concluded that the $\sigma$
undergoes a transition from being a bound state to a virtual bound state
somewhere between these two values of the pion mass. This is in agreement
with the model results displayed in FIGs.~\ref{bve} and \ref{bvk},
with the transition from bound state to virtual state occurring at
$m_\pi=292.35$~MeV. Also the typical $S$-wave behavior of the $\sigma$ pole
near threshold as observed in Ref.~\cite{Rodas23a} is in conformity with our
findings above. Still regarding a real $\sigma$ pole, the conclusion in
the other very recent LQCD paper \cite{Rodas23b} that the $\sigma$ appears to
pass through a narrow virtual-state region upon transitioning from a bound
state to a (subthreshold) resonance is qualitatively what one also observes in
FIGs.~\ref{bve} and \ref{bvk}.

Finally we compare the model's $\sigma$-pole resonance trajectories in
FIGs.~\ref{rese} and \ref{resk} with the LQCD results in
Refs.~\cite{Briceno17,Rodas23b}.
As mentioned before, the former lattice computation for $m_\pi=236$~MeV
resulted in complex pole positions strongly varying with the employed
parametrizations of the computed real amplitudes, such as $K$-matrix,
relativistic Breit-Wigner, and other ansatzes, with or
without an Adler zero. From FIG.~5 in Ref.~\cite{Briceno17}, I extract
an energy range of about 590--760~MeV for the extracted poles' real parts and
approximately 280--460~MeV for the widths, i.e., twice the modulus of the
imaginary parts. On top of that are the quite large error bars, up to
roughly $\pm80$~MeV in one case. So even accounting for the uncertainties in
the various resonance pole positions, they all lie well above the $\pi\pi$
threshold at 472~MeV. In this respect, it is interesting to note that two of
the lowest mass predictions, albeit still above 500~MeV, concern
parametrization with an Adler zero. Namely, in Ref.~\cite{Bugg03} a
relativistic Breit-Wigner form with an explicit Adler zero in the $s$-dependent
width was used to fit $S$-wave $\pi\pi$ phase shifts, which in combination
with other data yielded a $\sigma(500)$ pole position with real part
$(533\pm25)$~MeV. Also, in Ref.~\cite{BR21} a very simple yet unitary
Breit-Wigner form was shown to overestimate the resonance mass. In comparison,
the present unitary and analytic model predicts a very small interval
460--463~MeV for the real part of the $\sigma(500)$ pole while covering a 
wide range of pion masses, including $m_\pi=236$~MeV. Nevertheless, a
significant LQCD improvement was accomplished in Ref.~\cite{Rodas23b} by
imposing constraints through dispersive approaches, with the corresponding
resonance pole positions displayed in FIG.~3 of that paper. Note that the pion
mass in this case was taken at 239-MeV, which owing to an improved extraction
\cite{Rodas23a} corresponds to $m_\pi=236$~MeV used in Ref.~\cite{Briceno17}.
From the figure I now
extract the ranges 498--586~MeV for the real parts and 394--506~MeV for the
widths, with the largest error in the real part being $\pm82$~MeV. These
values allow for a possible subthreshold $\sigma$ resonance at
$m_\pi=239$~MeV for two of the employed dispersive methods, as found in the
present model (see FIGs.~\ref{rese} and \ref{resk}). The largest discrepancy
between lattice and model we observe for the width of the $\sigma$, which
is at least 350~MeV in Ref.~\cite{Rodas23b} and only about 220~MeV in the
model for $m_\pi=236$~MeV. So I conclude that it would be worthwhile to 
compute the $\sigma$ resonance on the lattice for some pion masses between
239~MeV and 283~MeV, in order to see how the width can
get down from about 400--500~MeV to zero very fast over a relatively small
range of pion masses, namely of the order of 40--50~MeV.


\begin{thebibliography}{99}

\bibitem{PDG22}
R.~L.~Workman {\it et al.} [Particle Data Group],
PTEP {\bf2022}, 083C01 (2022).

\bibitem{RB18}
G.~Rupp and E.~van Beveren,
Acta Phys.\ Polon.\ B Supp.\ {\bf11}, 455 (2018)
[arXiv:1806.00364 [hep-ph]].

\bibitem{Jaffe77}
R.~L.~Jaffe,
Phys.\ Rev.\ D {\bf15}, 267 (1977).

\bibitem{RB17}
G.~Rupp and E.~van Beveren,
Chin.\ Phys.\ C {\bf41}, 053104 (2017) 
[arXiv:1611.00793 [hep-ph]].

\bibitem{Eef83}
C.~Dullemond, T.~A.~Rijken, E.~van Beveren, and G.~Rupp, 
{\it ``On the influence of hadronic decay on the properties of hadrons,''}
6th Warsaw Symposium on Elementary Particle
Physics, 30 May -- 3 June 1983, Kazimierz, Poland,
Nijmegen report THEF-NYM-83.09.

\bibitem{Eef84}
E.~van Beveren, T.~A.~Rij\-ken, C.~Dullemond, and G.~Rupp,
Lect.\ Notes Phys.\ {\bf211}, 331 (1984).

\bibitem{Beveren83}
E.~van Beveren, G.~Rupp, T.~A.~Rijken, and C.~Dullemond,
Phys.\ Rev.\ D {\bf27}, 1527 (1983)

\bibitem{Beveren86}
E.~van Beveren, T.~A.~Rij\-ken, K.~Metzger, C.~Dullemond, G.~Rupp, and
J.~E.~Ribeiro,
Z.\ Phys.\ C {\bf30}, 615 (1986)
[arXiv:0710.4067 [hep-ph]].

\bibitem{Briceno18-1}
R.~A.~Briceno, J.~J.~Dudek, and R.~D.~Young,
Rev.\ Mod.\ Phys.\ {\bf90}, 025001 (2018) 
[arXiv:1706.06223 [hep-lat]].
 
\bibitem{Briceno17}
R.A.~Briceno, J.J.~Dudek, R.G.~Edwards, and D.J.~Wilson,
Phys.\ Rev.\ Lett.\ {\bf118}, 022002 (2017)
[arXiv:1607.05900 [hep-ph]].

\bibitem{Briceno18-2}
R.~A.~Briceno, J.~J.~Dudek, R.~G.~Edwards, and D.~J.~Wilson,
Phys.\ Rev.\ D {\bf97}, 054513 (2018)
[arXiv:1708.06667 [hep-lat]].

\bibitem{BR03}
E.~van Beveren and G.~Rupp,
Int.\ J.\ Theor.\ Phys.\ Group Theor.\ Nonlin.\ Opt.\ {\bf11},
179 (2006)
[arXiv:hep-ph/0304105].

\bibitem{BR09}
E.~van Beveren and G.~Rupp,
Annals Phys.\ {\bf324}, 1620 (2009)
[arXiv:0809.1149 [hep-ph]].

\bibitem{Rodas23a}
A.~Rodas {\it et al.} [Hadron Spectrum Collaboration],
Phys.\ Rev.\ D \/{\bf108}, 034513 (2023)
[arXiv:2303.10701 [hep-lat]].

\bibitem{Rodas23b}
A.~Rodas, J.~J.~Dudek, and R.~G.~Edwards,
arXiv:2304.03762 [hep-lat].

\bibitem{Gao22}
X.~L.~Gao, Z.~H.~Guo, Z.~Xiao, and Z.~Y.~Zhou,
Phys.\ Rev.\ D {\bf105}, 094002 (2022)
[arXiv:2202.03124 [hep-ph]].

\bibitem{BR23}
E.~van Beveren and G.~Rupp,
Phys.\ Rev.\ D {\bf107}, 058501 (2023)
[arXiv:2202.08809 [hep-ph]].

\bibitem{BR21}
E.~van Beveren and G.~Rupp,
World Scientific, Gribov-90 Memorial Volume, pp. 201--216 (2021)
[arXiv:2012.04994 [hep-ph]].

\bibitem{Micu69}
L.~Micu,
Nucl.\ Phys.\ B {\bf 10}, 521 (1969).

\bibitem{Bali05}
G.~S.~Bali {\it et al.} [SESAM Collaboration],
Phys.\ Rev.\ D {\bf71}, 114513 (2005)
[arXiv:hep-lat/0505012].

\bibitem{B82}
E.~van Beveren,
Z.\ Phys.\ C {\bf17}, 135 (1983)
[arXiv:hep-ph/0602248].

\bibitem{BR99a}
E.~van Beveren and G.~Rupp,
Eur.\ Phys.\ J.\ C {\bf10}, 469 (1999)
[arXiv:hep-ph/9806246].

\bibitem{BR99b}
E.~van Beveren and G.~Rupp,
Eur.\ Phys.\ J.\ C {\bf11}, 717 (1999)
[arXiv:hep-ph/9806248].

\bibitem{BR99c}
E.~van Beveren and G.~Rupp,
Phys.\ Lett.\ B {\bf454}, 165 (1999)
[arXiv:hep-ph/9902301].

\bibitem{BR20}
E.~van Beveren and G.~Rupp,
Prog.\ Part.\ Nucl.\ Phys.\ {\bf117}, 103845 (2021)
[arXiv:2012.03693 [hep-ph]].

\bibitem{Bugg06}
D.~V.~Bugg, private communication (2006).

\bibitem{BBKR06}
E.~van Beveren, D.~V.~Bugg, F.~Kleefeld, and G.~Rupp,
Phys.\ Lett.\ B {\bf641}, 265 (2006)
[arXiv:hep-ph/0606022].

\bibitem{minuit}
https://root.cern.ch/download/minuit.pdf

\bibitem{Hanhart14}
C.~Hanhart, J.~R.~Pelaez, and G.~Rios,
Phys.\ Lett.\ B {\bf739}, 375 (2014)
[arXiv:1407.7452 [hep-ph]].

\bibitem{Albaladejo12}
M.~Albaladejo and J.~A.~Oller,
Phys.\ Rev.\ D {\bf86}, 034003 (2012)
[arXiv:1205.6606 [hep-ph]].

\bibitem{Bugg03}
D.~V.~Bugg,
Phys.\ Lett.\ B {\bf572}, 1 (2003)
[Erratum-ibid {\bf595}, 556 (2004)].

\end{thebibliography}
\end{document}